\documentclass[twocolumn,showpacs,preprintnumbers,amsmath,amssymb]{revtex4-1}

\usepackage{graphicx}
\usepackage{graphics}
\usepackage{dcolumn}
\usepackage{bm}
\usepackage{epstopdf}
\usepackage{mathrsfs}
\usepackage{amssymb}
\usepackage{amsmath}
\usepackage[pdftex]{hyperref}
\usepackage{natbib}

\newcommand{\beq}{\begin{eqnarray}}
\newcommand{\eeq}{\end{eqnarray}}
\newcommand{\be}{\begin{eqnarray*}}
\newcommand{\ee}{\end{eqnarray*}}

\begin{document}

\preprint{CERN-PH-TH/2012-291}
\preprint{LU-TP 12-37}

\title{New picture of jet quenching dictated by color coherence}

\author{Jorge Casalderrey-Solana}
\affiliation{Departament de d'Estructura i Constituents de la Mat\`eria 
and Institut de Ci\`encies del Cosmos (IC-CUB),\\
Universitat de Barcelona, Mart\'i i Franqu\`es 1, 08028 Barcelona, Spain}

\author{Yacine Mehtar-Tani}
\affiliation{Institut de Physique Th\'eorique,
CEA Saclay, F-91191 Gif-sur-Yvette, France}

\author{Carlos A. Salgado}
\affiliation{Dep. de F\'isica de Part\'iculas,
U. de Santiago de Compostela,  E-15782 Santiago de Compostela, 
Galicia-Spain \\
and Physics Department, Theory Unit, CERN, CH-1211 Gen\`eve, Switzerland
}

\author{Konrad Tywoniuk}
\affiliation{Department of Astronomy and Theoretical Physics,
Lund University, S\"olvegatan 14A, SE-22 362 Lund, Sweden}

\date{\today}

\begin{abstract}
We propose a new description of the jet quenching phenomenon observed in nuclear collisions at high energies in which coherent parton branching plays a central role. This picture is based on the appearance of a dynamically generated scale, the jet resolution scale, which controls the transverse resolution power of the medium to simultaneously propagating color probes. Since from the point of view of the medium all partonic jet fragments within this transverse distance act coherently as a single emitter, this scale allows us to rearrange the jet shower into effective emitters. We observe that in the kinematic regime of the LHC, the corresponding characteristic angle is comparable to the typical opening angle of high energy jets such that most of the jet energy is contained within a non-resolvable color coherent inner core. Thus, a sizable fraction of the jets are unresolved, losing energy as a single parton without modifications of their intra-jet structure. This new picture provides a consistent understanding of the present data on reconstructed jet observables and constitute the basis for future developments. 
\end{abstract}

\maketitle

The phenomenon of ``jet quenching", in other words the modifications of the structure of jets in a heavy-ion environment, is one of the main tools  to determine the properties of QCD matter under extreme conditions. The suppression of high transverse momentum particles observed in Au+Au collisions at RHIC stands out as one of its key discoveries \cite{Adcox:2004mh,*Back:2004je,*Arsene:2004fa,*Adams:2005dq}. A similarly strong suppression is  observed in Pb+Pb collisions at the LHC \cite{Aamodt:2010jd,*CMS:2012aa}. Reconstructed jet observables, which hold the promise to be extremely versatile probes for an unprecedented characterisation of the medium, have recently been measured as well \cite{Chatrchyan:2012gw,*:2012is}. Clearly, the success of this program relies crucially on a detailed understanding of the interaction of jets with QCD matter. However, the approaches successfully applied at RHIC need important refinements to describe the sub-leading structure of the jet, treated so far in an oversimplified manner. This calls for a complete theory of jets in a medium.

The perturbative QCD description of jets in the vacuum is built up of partial information from limiting cases where first-principle calculations are possible and improves systematically when increasing precision is needed. A similar approach is being followed for the medium case. One of the essential features of the vacuum branching process is color coherence in multi-gluon radiation, a problem which has recently been addressed for the medium in a series of papers  \cite{MehtarTani:2010ma,*MehtarTani:2011tz,*MehtarTani:2011jw,*CasalderreySolana:2011rz,*Armesto:2011ir,*MehtarTani:2011gf,*MehtarTani:2012cy}. The results obtained in these studies of the {\it antenna radiation}  will be used here to put forward a new and appealing picture of the  problem of in-medium jet evolution.

The underlying physical picture arising from these studies is simple: for medium-induced gluon radiation, a jet is composed of a set of colored emitters, which may not correspond to the actual number of partons in the shower. A given medium configuration defines an effective number of emitters off of which induced radiation takes place, while inside each of these effective sources the angular-ordered vacuum radiation occurs. The resolution defining these emitters is determined dynamically by the medium properties and not by the kinematics of the radiated gluon, as for the vacuum. A key observation is that the medium gives rise to a transverse resolution scale which determines the role of the coherence effects: partons separated in transverse plane less than this characteristic size remain color correlated and hence emit induced gluons as a single parton. 

Two generic properties for the in-medium jet evolution are direct consequences of this effect: i) the total energy loss of the jet is smaller than in the case of totally incoherent emission off each of its constituent partons; ii) each of the effective emitters fragments as in vacuum. In particular, the leading particles of the jet define a coherent ``inner core" of the reconstructed jet, shielded by coherence against medium modifications of its structure and only loosing energy by induced radiation as a single parton. As will be shown below, for typical LHC kinematics there is a significant probability that the experimentally reconstructed jet with cone parameter $R$ accommodates only one resolved charge which contains the leading constituents carrying nearly all of the total jet transverse energy. 
\begin{figure}
\centering
\includegraphics[width=0.5\textwidth]{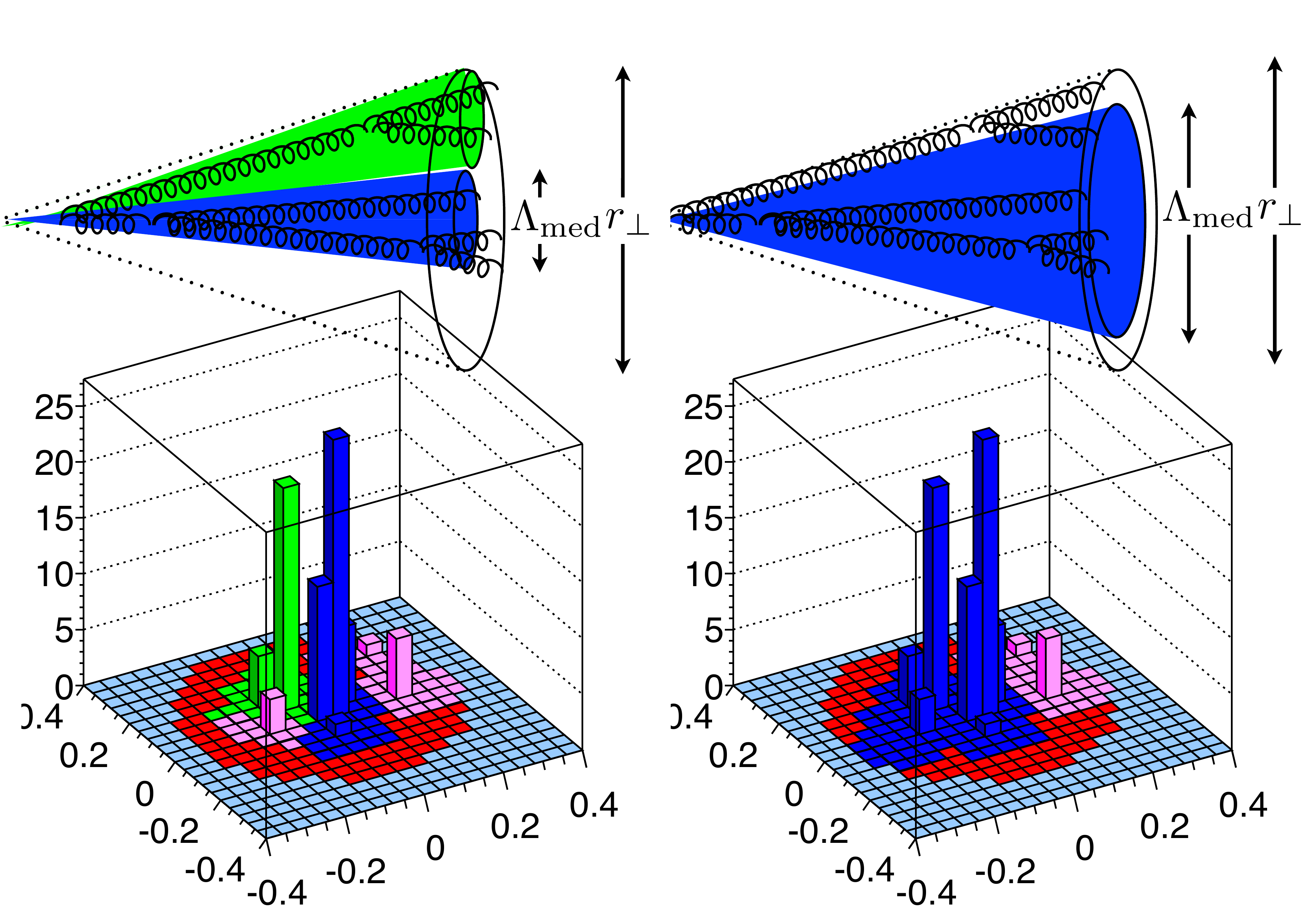}
\caption{A sample jet event resolved with $R_\text{med} = 0.1$ (left panel) and $0.15$ (right panel). The blue histogram denotes the hardest resolved sub-jet, the green the next-to-hardest one, while the pink histogram denotes soft fragments.}
\label{fig:JetScales}
\end{figure}

{\bf From the antenna to the jet}.
The dynamics of a QCD jet in vacuum is described in terms of the scales of the problem. The initial hardness, given by the jet transverse mass $E\Theta_{\rm jet}$, where $E$ is the jet energy and $\Theta_{\rm jet}$ its aperture, is distributed among several constituents in the course of a branching process. Multiple emissions in the shower are governed by color coherence which can most easily be understood in the context of the {\it antenna radiation}, the soft gluon radiation off a pair of highly energetic color correlated partons. The antenna  serves as the building block for a probabilistic scheme of jet evolution.

In the radiation process from any such antenna of opening angle $\Theta$, the emitted gluon transverse wavelength $\lambda_\perp$, which is related to its transverse momentum by $\lambda_\perp \sim 1/k_\perp$, needs to be compared to the transverse separation of the pair at the time of formation of the gluon, $r_\perp=\Theta \,t_\text{f}$, with  $t_\text{f}\sim k^2_\perp/\omega$ and $\omega$ the gluon frequency . If $\lambda_\perp > r_\perp$, the gluon cannot resolve the two components of the antenna which act coherently as a single emitter; in the opposite case, when $\lambda_\perp < r_\perp$, the radiative spectrum is the superposition of independent gluon emissions off each of the antenna components. In other words, radiation with $\lambda_\perp > r_\perp$ is only sensitive to the total charge. This relation takes a particularly simple form for the angular distribution of gluons, namely gluons emitted at small angles $\theta<\Theta$ resolve the individual charges while those with $\theta>\Theta$ behave as if emitted off the total charge. This generic feature is responsible for the angular ordering constraint \cite{Dokshitzer:1991wu}.

The presence of a deconfined medium introduces a new transverse length scale into the problem, which we simply denote by $\Lambda_\text{med}$, defining the  transverse size of the color correlations of the plasma as seen by a probe. The response of a single, energetic parton immersed in this environment is the radiation of modes with $k_\perp \lesssim 1/\Lambda_\text{med}$, giving rise to an energy depletion of the projectile. The nature of this radiation has been extensively discussed in the literature and is generically referred to as the BDMPS-Z spectrum \cite{Baier:1996kr,*Baier:1996sk,*Zakharov:1996fv,*Zakharov:1997uu,*Wiedemann:2000za}. For more than one simultaneously propagating parton, this medium-induced component will also be accompanied by a modification of the color correlation structure among the different charges \cite{MehtarTani:2010ma,*MehtarTani:2011tz,*MehtarTani:2011jw,*CasalderreySolana:2011rz,*Armesto:2011ir,*MehtarTani:2011gf,*MehtarTani:2012cy}, which we proceed to discuss. 

Let us start by the simplest case of a single antenna in a static and homogeneous medium of length $L$. The maximal degree of decoherence, due to color randomization, of the two constituents of the antenna is controlled by \cite{MehtarTani:2010ma,*MehtarTani:2011tz,*MehtarTani:2011jw,*CasalderreySolana:2011rz,*Armesto:2011ir,*MehtarTani:2011gf,*MehtarTani:2012cy}
\beq
\label{eq:Delta}
\Delta_{\text{med}}\simeq 1- {\rm e}^{-\frac{1}{12}\hat qLr_\perp^2} \equiv1-{\rm e}^{-\left(\Theta/\theta_c\right)^2}\, .
\label{eq:dipoleqhat}
\eeq
Here $\hat q$ is the well known quenching parameter, characterizing the degree of momentum broadening in the transverse plane per unit length, and $r_\perp=\Theta L$. Moreover, $1/\Lambda_\text{med}^2\equiv\hat qL$. Since the first jet splitting defines the largest antenna in the jet, it is now simple to discuss the two possible scenarios, depicted in Fig. \ref{fig:JetScales}, for a jet with  opening angle $\Theta=\Theta_{\rm jet}$.

When $\Theta_\text{jet} \ll \theta_c$, the whole jet is not resolved by the medium. Therefore, all its components act as a single emitter. This gives rise to two central consequences. Firstly, the fragmentation pattern of the jet is unmodified compared to the vacuum. Secondly, the jet energy is depleted coherently proportionally to the color charge of the jet initiator (e.g., with color charge $C_R = C_F$ in the case of a quark jet). In other words, for a jet energy loss $\Delta E$, each parton reduces its energy by a constant factor $1-\Delta E/E$. This is a manifestation of color transparency for highly collimated jets.

For the case $\Theta_\text{jet} \gg \theta_c$, on the other hand, some parts of the jet can be resolved by the medium depending on the formation time of the different jet fragments.  Nevertheless, the partons within the jet may be reorganized into a reduced {\it effective number of emitters} which are sensitive to medium effects in the shower. 

{\bf An estimate of the relevance of color coherence for LHC conditions.} 
\begin{figure}
\includegraphics[width=0.45\textwidth]{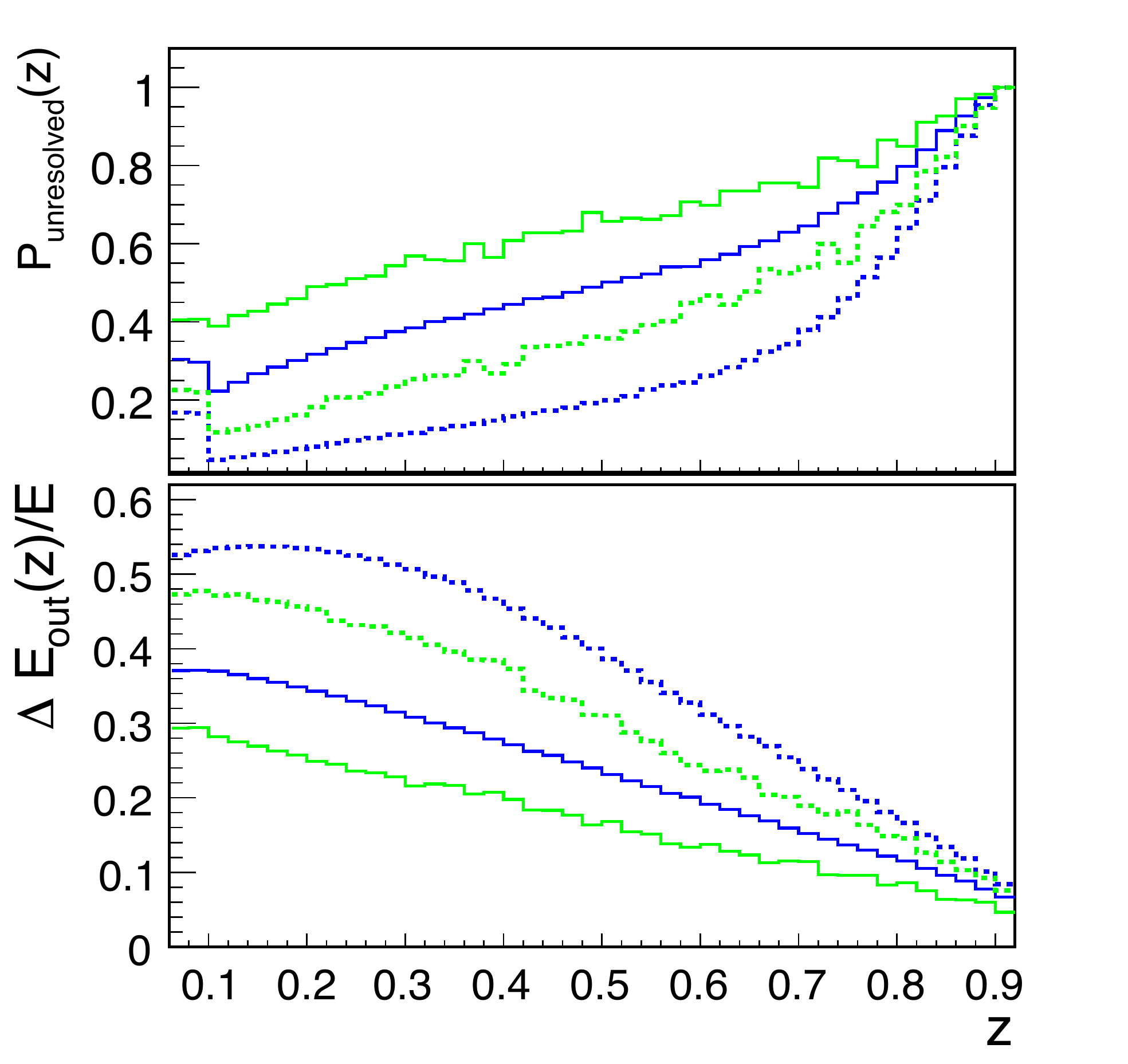}\\
\caption{The probability of not resolving the substructure of the hardest jet in the event at $R_{\rm med}$ (upper panel) and the energy missing from the leading sub-jet (lower panel) for jets traversing a medium with $K=1,10$ for continuous and dashed lines. The entries are trimmed ($f_{cut} = 0.1$) anti-k$_t$ jets with $R = 0.2$ in $p_T$-bins of 100-150 GeV (blue)  and 200-300 GeV (green).}
\label{fig:unresolved}
\end{figure}
As a proof-of-principle study,  we have analyzed the transverse structure of vacuum jet showers in the kinematic range of the LHC. Using  PYTHIA 8.150 \cite{Sjostrand:2007gs}, we studied jet events at partonic level in  p+p collisions at 2.76 TeV  identified via the anti-k$_t$ algorithm, as implemented in FastJet 3.0.3 \cite{Cacciari:2005hq,*Cacciari:2011ma}. Since the resolution power of the medium depends upon the geometry encountered by the jet, we have embedded these events into an evolution model for the plasma. Each event was assigned a production point in the transverse plane according to the $N_\text{coll}$ distribution in the Glauber model of a Pb+Pb collision at the same energy and a random direction. The time-dependent energy density of the quark-gluon plasma, $\varepsilon$, along the jet trajectory was sampled from a 3D hydrodynamical code \cite{Hirano:2010je,*Hirano:2010jg} enabling us to define $\hat q(\xi) = 2K\epsilon(\xi)^{3/4}$. This procedures assigns to each jet a resolution scale $R_{\rm med}$, which equals $\theta_c$ computed for this jet trajectory sampling a local $\hat q(\xi)$ along its path, and is given by
\beq
\label{eq:thetaccomplicated}
R_{\rm med}\equiv 2 \left(\int d\xi \,\xi^2\hat q(\xi)\right)^{-1/2} \,,
\eeq
see Eq. (\ref{eq:dipoleqhat}). To identify what jet substructure would be resolved by the medium, we have performed a re-clustering of the jet constituents using the Cambridge/Aachen algorithm with $R_\text{med}$ as the resolution scale. Examples of this procedure can be seen in Fig. \ref{fig:JetScales}. As a first estimate of the overall uncertainties related to the soft constituents of the jet, we have discarded substructures below a fractional transverse momentum cut-off  $f_{cut} = 0.1$. This provides our estimate of the number of sub-jet structures which the medium can resolve.

For this jet sample we compute the probability that the jet is unresolved, i.e. that it contains only one sub-jet of size $R_{\rm med}$ inside the jet reconstructed with radius $R = 0.2$, as the function of the partonic longitudinal momentum fraction $z$, see the upper panel in Fig.~\ref{fig:unresolved}. This probability of unresolved jets is large and clearly decreases when the resolution power of the medium increases (with increasing $K$). As expected, we observe that the presence of hard fragments in the jet are correlated with very collimated jet structures which are mostly not resolved by the medium. We have also checked that this probability is only mildly dependent on the jet reconstruction radius $R$, since increasing the jet radius tends to collect soft jet fragments which do not contribute much to the energy balance. This is also clear by considering the fraction of the total jet energy missing from this leading sub-jet, again as a function of the parton constituent $z$, see lower panel of Fig.~\ref{fig:unresolved}. We have explicitly checked that this quantity does not depend on the details of the substructure analysis. 

The above estimates clearly demonstrate the relevance of color coherence effects for jets in heavy-ion collisions at the LHC which leads to a reduction of the number of effective emitters for medium-induced radiation in the parton shower. We have put emphasis on the simplest possible situation of only one leading sub-jet and showed the large probability of this configuration. This implies  not only a smaller energy loss with respect to the totally incoherent case but also that many  of the jets in a heavy ion environment follow a (angular ordered) vacuum fragmentation process, since all their fragments are contained within one unresolved emitter. It is worth pointing out that our estimates should be taken as conservative since we have not taken into account the effect of energy loss in the studied jet distributions. Would this effect have been taken into account, the number of resolved jets for a given jet energy would obviously decrease, as a result of the bias due to the steeply falling jet spectrum, since they loose more energy.

{\bf A new picture of in-medium parton shower.} 
For the remaining sample of jets with more than one emitter the role of the subleading sub-jets cannot be neglected for a precise determination of medium effects. On general grounds, in these cases coherence effects should be taken into account successively in the parton branching. These aspects are completely novel to the modeling of ``jet quenching" and call for a rigorous formulation which is beyond the scope of this Letter.  Nevertheless, based on the discussion of the antenna spectrum, we can postulate a more detailed picture of a parton shower including coherent branching in a medium. This picture is grounded on a study of the  relevant scales appearing  at each individual branching, analogous to the one summarizing the vacuum fragmentation pattern, as described previously. As in that case, the emission dynamics depend on the transverse separation of the partons in the antenna at the time of formation of the emitted fragment, $r_\perp\sim \Theta \,t_\text{f} $. For those emissions that give rise to additional in-medium antennas, $t_\text{f}<L$, the relevant decoherence parameter, $\Delta_{\text{med}}$, is obtained by replacing L with $t_\text{f}$ in Eq.~\eqref{eq:dipoleqhat} and $ \Lambda_\text{med}\sim 1/\sqrt{\hat q t_\text{f}}\,$.

When $r_\perp\ll \Lambda_\text{med}$, the antenna is smaller than the medium correlation length such that $\Delta_\text{med} \to 0$ and the emitted parton is formed before the medium resolves the antenna constituents. This immediately implies that medium-induced gluons, with $k_\perp \lesssim 1/\Lambda_\text{med}$ can only be produced coherently by the pair. Radiation not induced by the medium is also possible but due to the angular ordering restriction, it is constrained to $\lambda_\perp < r_\perp$ and  therefore to $k_\perp \gg 1/\Lambda_\text{med}$. In the opposite case, $r_\perp \gg \Lambda_\text{med}$, all interferences are cancelled by $\Delta_\text{med} \to 1$ \cite{MehtarTani:2010ma,*MehtarTani:2011tz,*MehtarTani:2011jw,*CasalderreySolana:2011rz,*Armesto:2011ir,*MehtarTani:2011gf,*MehtarTani:2012cy}. Since the medium resolves the individual emitters prior to the emission, the induced radiation is that off two independent color charges. Additionally, non-induced radiation is again possible not only for $\lambda_\perp < \Lambda_\text{med}$, as in the previous case, but also for $\lambda_\perp > \Lambda_\text{med}$ due to the decoherence of the pair. The latter possibility replaces the strict angular ordering and was called antiangular ordering in \cite{MehtarTani:2010ma,*MehtarTani:2011tz,*MehtarTani:2011jw,*CasalderreySolana:2011rz,*Armesto:2011ir,*MehtarTani:2011gf,*MehtarTani:2012cy}. 
Therefore, the antennas arising from hard modes inside the medium $k_\perp > 1/\Lambda_\text{med}$ remain unresolved by the medium while those originating from softer ones, $k_\perp < 1/\Lambda_\text{med}$  are resolved. In particular the hard modes $\lambda_\perp < \Lambda_\text{med}$ remain unresolved and create an effective `core' which fragments as in vacuum. 

This picture allows us to provide an estimate of the importance of coherence effects for different fragment momenta. According to angular ordering  gluons found at angles $\theta < \theta_c$, have transverse momenta $Q_0 < k_\perp < z E \theta_c$, where $Q_0$ is a regularization parameter. In other words, all fragments with $z>Q_0/E\theta_c$ will follow angular ordered, vacuum, fragmentation. Furthermore, the gluons with energies $\hat q\, L^2 < \omega < E$ will have $t_\text{f}<L$ so we can estimate that, as long as $E \Theta_{jet} \gg \hat q \,L^2 r_\perp$, a large fraction of the unresolved splittings  take place inside the plasma. Gluon emission in plasma with $\omega < \hat qL^2$ can be induced by the medium. This radiation occurs at large angles, $\theta > \theta_c$, and fully decoheres from the other jet constituents \cite{Blaizot:2012fh}. In addition, one should naturally expect large-angle vacuum radiation which gives rise to new effective charges, or sub-jets, which will replicate the same features as the jet `core' discussed above.

{\bf In summary}, this exploratory study have highlighted the role of the resolution power that a deconfined QCD medium has to multiple partonic color charges, originally produced in a correlated state as a result of the fragmentation of a high-energy jet, propagating through it. As we have demonstrated, the ability of the medium to differentiate the fragments in the jet crucially depends on whether the medium can break the color coherence among them; if the color coherence of the system is preserved, then it behaves as a single, effective charge in the medium. This effect, identified in the computations of the {\it antenna radiation} \cite{MehtarTani:2010ma,*MehtarTani:2011tz,*MehtarTani:2011jw,*CasalderreySolana:2011rz,*Armesto:2011ir,*MehtarTani:2011gf,*MehtarTani:2012cy}, allows to define an effective number of emitters for medium-induced gluon radiation which in general is much smaller than the number of partons in a vacuum parton shower. As was verified by numerical study, we have shown that a sizable part of the jets in the experimental conditions are unresolved by the medium, i.e. they act coherently as a single parton, specially those containing harder jet fragments. This clearly proves the relevance of color coherence in the understanding of the present data on jet quenching in heavy-ion collisions, opening new possibilities to the characterization of medium properties, with jets providing the resolution power for different length scales, achievable with increasing precision in the jet measurements. \\

\begin{acknowledgments}
 JCS is supported by a Ram\'on y Cajal fellowship and  by the research grants FPA2010-20807, 2009SGR502 and by the Consolider CPAN project. 
YMT is supported by the European Research Council under the Advanced Investigator Grant ERC-AD-267258.
CAS  is supported by European Research Council grant HotLHC ERC-2011-StG-279579, by Ministerio de Ciencia e Innovaci\'on of Spain under grant FPA2009-06867-E and by Xunta de Galicia. 
KT is supported  by the Swedish Research Council (contract number 621-2010-3326).
\end{acknowledgments}

\bibliographystyle{apsrev4-1}
\bibliography{cmst_jets_2.bib}
\end{document}